# Optimization of magnetic contrast layer for neutron reflectometry

A. Zubayer[1], F. Eriksson[1], N. Ghafoor[1], J. Stahn[2], J. Birch[1], A. Glavic[2]

1. Department of Physics, IFM, Linköping University, SE-581 83 Linköping, Sweden
2. PSI Center for Neutron and Muon Sciences, Forschungsstrasse 111, 5232 Villigen PSI, Switzerland

## Abstract

Neutron reflectivity is a powerful technique for probing density profiles in films, with applications across Physics, Chemistry, and Biology. However, challenges arise when dealing with samples characterized by high roughness, unknown scattering length density (SLD) with low contrast, very thin layers, or complex multi-layered structures, that cannot be uniquely resolved due to the phase problem. Incorporating a magnetic reference layer (MRL) and using polarized neutron reflectivity improves sensitivity and modeling accuracy by providing complementary information. In this study, we introduce a quantitative way to compare MRL systems in a model-free way. We apply this approach to demonstrate that CoTi alloys offer a superior solution as an MRL compared to the commonly used Fe or Ni-based MRLs. The low nuclear and magnetic scattering length densities of CoTi significantly enhance sensitivity, making it particularly advantageous for soft matter research. Furthermore, the tunable Co vs Ti ratio allows for optimization of the SLDs to achieve maximum sensitivity, establishing CoTi as a highly effective choice for MRL applications. The applied simulation framework for optimizing MRL sensitivity to a specific materials system and research question is a generic approach that can be used prior to growing the MRL for a given experiment.

## Introduction

Neutron Reflectometry (NR) is a versatile technique for analyzing the depth profile of thin films and layered materials, i.e. their thickness, density, and roughness. It's particularly impactful in soft matter research, where it probes complex systems like polymers, lipids, and biological macromolecules under near-natural conditions.[1–5] NR's ability to investigate interfaces and thin films in soft materials reveals crucial insights into phenomena such as polymer dynamics, protein adsorption, and lipid membrane organization.[2,6–8] Its sensitivity to water is also key for studying hydration effects in biological and soft matter contexts, making NR essential for advancements in material science and biotechnology. However, when it comes to studying non-magnetic layers with unknown scattering length density (SLD) and morphology, traditional NR encounters significant challenges due to the phase problem.[9–11] A low contrast between layer and substrate results in weak oscillations and leads to reduced sensitivity in the data analysis. Missing knowledge about the expected layer morphology and SLD profile can yield different models reproducing the measurements quite well, but with vastly different physical interpretation.[12–14] In experiments where one component of the sample is submerged in water this problem is often tackled by measuring different $H_2O$/$D_2O$ mixtures leading to contrast variation{references}. While this technique is proven it cannot always be applied as, for example, the change of contrast might require creation of a new sample that could have different SLD profile.

The introduction of magnetic reference layers (MRL) and use of Polarized Neutron Reflectivity (PNR) measurements addresses the challenges in a similar way. By exploiting the distinct SLD contrasts for spin-up and spin-down neutrons, the use of an MRL then allows for the generation of two separate reflectivity curves. This dual-curve approach can lift ambiguity and help to overcome the phase problem, significantly improving the sensitivity and thus the reliability of data interpretation. It effectively circumvents the ambiguity associated with weak contrast by the fact that at least one of the spin states cannot have the same SLD as the substrate. Thus, while NR is quicker and more accessible, PNR with MRL offers the capability to reveal features that NR might overlook. It also provides greater reliability due to its higher sensitivity and reduced risk of misinterpretation.

A current commonly applied MRL has been, for example, an iron layer characterized by a nuclear SLD of approximately $8 \cdot 10^{-6}$ Å$^{-2}$ and a magnetic SLD of about $5 \cdot 10^{-6}$ Å$^{-2}$.[15–18] Consequently, the SLD for spin-up neutrons reaches about $13 \cdot 10^{-6}$ Å$^{-2}$ (sum of nuclear and magnetic SLDs), while for spin-down neutrons, it drops to around $3 \cdot 10^{-6}$ Å$^{-2}$ (difference between nuclear and magnetic SLDs). At first glance, this difference might seem advantageous, ensuring that at least one of the SLDs offers significant contrast with the sample of interest. However, the significantly high SLD for spin-up neutrons in Fe leads to the predominant reflection of neutrons caused by Fe, overshadowing the reflective properties of the sample under investigation, particularly when its SLD ranges from 0 to $4 \cdot 10^{-6}$ Å$^{-2}$, as is common with organic materials.[19–21] The same overshadowing effect is seen in nickel based, low magnetic, MRLs. To prevent the MRL from dominating the signal and ensure the sample of interest remains detectable, it's crucial to maintain a low SLD for MRL across both spin-up and spin-down neutrons.

Following the approach outlined in[22], it is possible to tune both nuclear and magnetic SLDs to reach the desired contrasts. Co has the lowest nuclear SLD among the ferromagnetic elements, around $2.3 \cdot 10^{-6}$ Å$^{-2}$. However, it would be better to have an even lower nuclear SLD and perhaps even lower magnetic SLD so that the sum of nuclear and magnetic SLD is not above the SLD of the substrate. Thus, we suggest diluting Co with Ti which has a negative SLD of $-1.9 \cdot 10^{-6}$ Å$^{-2}$ to achieve a mixture of CoTi. The dilution of Co with Ti will also lead to a decrease in magnetic SLD so that the spin-up SLD is dampened preferably below the SLD of the sample of interest.

The sample (/layer) of interest (SOI) in this study represents a range of typical thin-film layers encountered in soft matter and related fields, with specific variations in thickness, roughness, and scattering length density (SLD). Important to note is that in this study we refer to the sample of interest to the layer of interest and not the entire sample put into the beam. A "normal" SOI is modelled as a 500 Å thick layer, which is representative of many soft matter systems such as polymer films, lipid bilayers, protein-coated surfaces or perovskite layers.[33–36] These materials often fall within this thickness range due to their structural stability and relevance in applications like membranes or thin coatings. The "rough" SOI maintains the same 500 Å thickness but features an increased roughness of 50 Å instead of the typical 15 Å. Conversely, some samples, especially in soft matter, can be very thin; our "thin" SOI represents such cases with a thickness of 50 Å and a moderate roughness of 15 Å, mimicking self-assembled monolayers, thin polymer brushes, or confined lipid bilayers.[37,38] The SLD of the SOI can vary between $1 \cdot 10^{-6}$ Å$^{-2}$ and $3 \cdot 10^{-6}$ Å$^{-2}$, reflecting the diverse range of soft matter compositions, from low-density polymers to denser organic materials. To capture these variations, this study uses a matrix of nine different SOIs, combining three structural types (normal, rough, and thin) with three different SLD values (1, 2, and $3 \cdot 10^{-6}$ Å$^{-2}$). This approach allows for a comprehensive investigation of how magnetic reference layers (MRLs) perform in terms of sensitivity to these diverse SOIs. The MRLs studied include a scenario with no MRL, a Fe-based MRL, a Ni-based MRL, and a novel CoTi MRL. These configurations are evaluated for their effectiveness in distinguishing the SOI's structural and magnetic properties. By systematically exploring the sensitivity of these MRLs across the nine SOIs, this paper provides insights into how well each MRL configuration performs under varying experimental conditions, highlighting the potential advantages of the CoTi MRL in delivering robust and reliable results for soft matter systems.

Table 1. Different samples of interest (SOIs) and their SLD ($\rho$), thickness ($\Lambda$) and surface roughness ($\sigma$).

| Type | SOI SLD ($10^{-6}$ Å$^{-2}$) | Thickness, $\Lambda$ (Å) | Roughness, $\sigma$ (Å) |
|---|---|---|---|
| **Normal** | 1 | 500 | 15 |
| **Normal** | 2 | 500 | 15 |
| Normal | 3 | 500 | 15 |
| Rough | 1 | 500 | 50 |
| Rough | 2 | 500 | 50 |
| Rough | 3 | 500 | 50 |
| Thin | 1 | 50 | 15 |
| Thin | 2 | 50 | 15 |
| Thin | 3 | 50 | 15 |

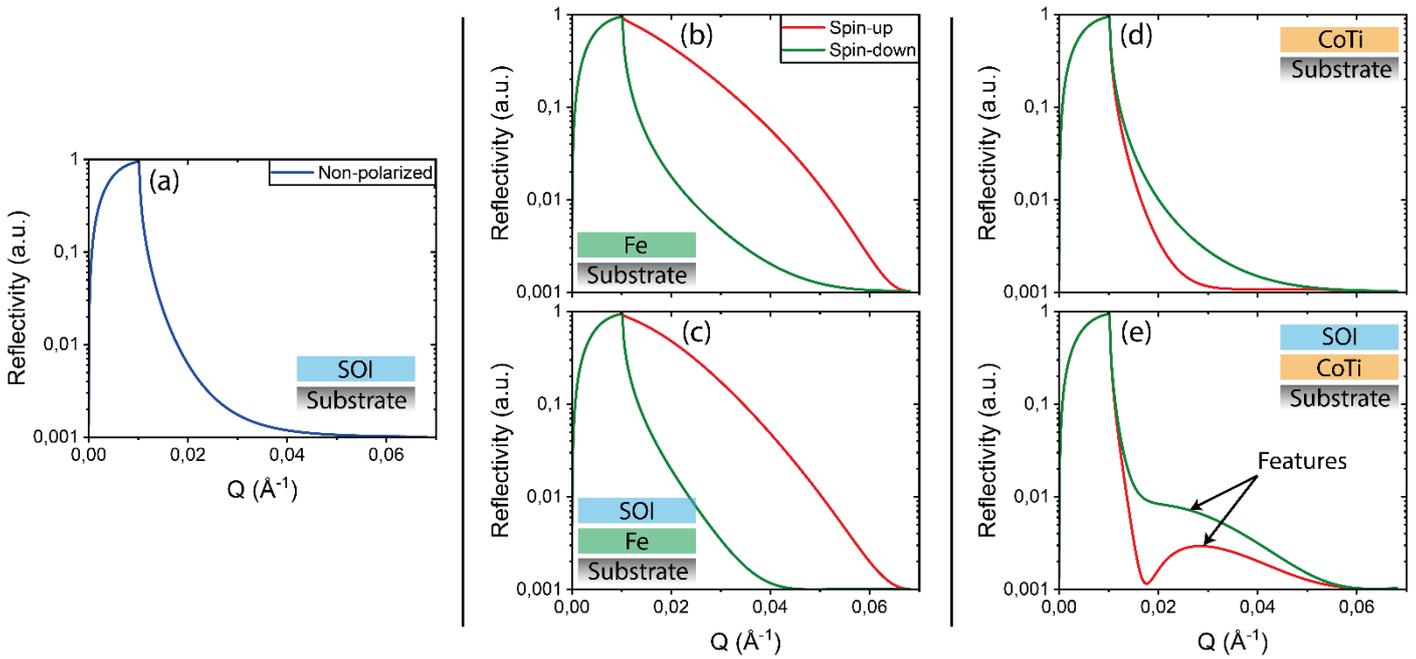

Figure 1. Polarized neutron reflectivity (PNR) simulations of 5 cases. Sample of interest (SOI) layer with an SLD of $2 \cdot 10^{-6}$ Å$^{-2}$ on substrate (a), Fe layer on substrate (b), SOI on Fe on substrate (c), Co$_{0.73}$Ti$_{0.27}$ layer on substrate (d) and SOI on Co$_{0.73}$Ti$_{0.27}$ on substrate (e). Where the substrate is Si for all samples. The simulations include a beam footprint and background of 0.001. Q stands for the reciprocal space vector.

As seen in Figure 1, only an SOI shows no features, while using an Fe MRL it is hard to distinguish between only MRL and MRL with SOI. The suggested CoTi MRL and MRL with SOI shows a clear difference as well as showing evident features of the SOI.

The figures of merit (FOMs) for polarized neutron reflectometry (PNR) that we developed quantify the effectiveness of magnetic reference layers (MRLs) in distinguishing spin-polarized interactions with samples of interest (SOIs). The sensitivity FOM (SFM) integrates the absolute sensitivity across the momentum transfer range, accounting for both spin-up and spin-down neutron channels. The magnetic contrast FOM (MCF) measures the difference between spin states, representing the information gain. Finally, the total sensitivity figure (TSF) combines the contributions from all sensitivity metrics across multiple SOI types, providing a comprehensive evaluation of the MRL's performance. Details can be found in the Methods section as well as in the Supplementary Information.

In this study, we use simulations to characterize the efficacy of MRLs and of the proposed CoTi MRLs against Fe or Ni-based MRLs and use experimental results to find the optimal ratio between Co and Ti for the best possible performance of MRL.

# Methods

The GenX3 software [30] was employed for conducting PNR simulations to analyze the reflective responses for both spin-up and spin-down neutrons across the samples listed in Table 1. Additionally, GenX3 was utilized to fit the data collected from PNR measurements. This software uses the Parratt recursion formalism, which recursively simulates and fits reflectivity data by considering each interface to compute the overall detected intensity. The fitting results generated by the software were instrumental in determining the parameters such as thicknesses, roughness, and both nuclear and magnetic SLDs. Lastly, GenX3 was also used to simulate the sensitivities of Fe, Ni and CoTi magnetic reference layers. A custom Python script was developed within GenX to calculate reflectivity and sensitivity curves for various magnetic reference layers (MRLs) and samples of interest (SOIs). Details on the simulations can be found in the Supplementary Information and the codes and files are available at: https://github.com/Azubayer/GenX-sensitivity.

Table 2. Different magnetic reference layers (MRLs) and their nuclear and magnetic SLD ($\rho$). Where all MRLs have a thickness, $\Lambda$ = 100 Å, and surface roughness, $\sigma$ = 5 Å.

| MRL | nSLD ($10^{-6}$ Å$^{-2}$) | mSLD ($10^{-6}$ Å$^{-2}$) |
|---|---|---|
| Fe | 8 | 5 |
| Ni | 9.4 | 1 |
| CoTi | 1 | 2 |

Table 2 shows the MRLs used for the reflectivity, scattering length density depth profiles and sensitivity simulations.

Specular reflectivity, $R(Q)$, represents the fraction of neutrons reflected from the sample as a function of the momentum transfer $Q = \frac{4\pi}{\lambda}\sin(\theta)$, where $\lambda$ is the neutron wavelength and $\theta$ is the angle of incidence.[29] A critical factor in this process is the scattering length density (SLD), $\rho$, which serves as a fundamental parameter describing the interaction of neutrons with the material.[30] The reflectivity is influenced by the variation in the scattering length density (SLD) across interfaces. Reflectivity is calculated using matrix formalism, which incorporates the nuclear ($\rho_n(z)$) and magnetic ($\rho_m(z)$) SLD profile throughout the sample structure. The algorithm also accounts for interference effects between neutron waves reflected at different interfaces within the multilayer structure, producing characteristic oscillations in the reflectivity $R(Q)$. Additionally, the impact of interface roughness is considered, as it introduces Gaussian damping, reducing the reflectivity intensity at higher Q-ranges. By modelling reflectivity across various candidate reference layers, their performance in extracting information about SOIs over specific Q-ranges can be assessed, enabling the MRL optimization for specific polarized neutron reflectometry experiments. In PNR, the spin-dependent SLD can be calculated from the nuclear and magnetic contributions:

$$\rho_\uparrow = \rho_n + \rho_m \qquad 1$$

$$\rho_\downarrow = \rho_n - \rho_m \qquad 2$$

With the nuclear SLD, $\rho_n = \sum_i N_i b_i$, where N is the atomic number density and b is the coherent scattering length, and the magnetic SLD[31,32], $\rho_m = (\hbar\gamma/2\mu_B)M$, where M represents the layer magnetization. For an optimal magnetic reference layer, the magnetic SLD $\rho_m$ must provide sufficient contrast with the nuclear SLD of adjacent layers. This contrast ensures that the spin polarization dependence of the reflectivity remains detectable over a broad Q-range, enabling accurate characterization of SOI structures. For optimal magnetic reference layers, the magnetic SLD $\rho_m$ should provide sufficient contrast with the nuclear SLD of the adjacent layers, ensuring that the spin polarization dependence of the reflectivity is detectable over a broad $Q$-range. The reflectivity in PNR experiments is measured for neutrons in two distinct polarization states: spin-up (↑) and spin-down (↓). These states interact differently with the magnetic field of the reference layer, leading to spin-dependent reflectivity curves.

By analyzing these curves, we can assess how well an SOI is measured using a reference layer. We define the sensitivity ($S_{\uparrow/\downarrow}(Q)$) to quantify how well an experiment can distinguish between the reference substrate and the additional SOI. It is defined for the spin-up and spin-down channels as:

$$S_{\uparrow/\downarrow}(Q) = \frac{R_{\uparrow/\downarrow}^{\text{sub}}(Q) - R_{\uparrow/\downarrow}^{+\text{SOI}}(Q)}{R_{\uparrow/\downarrow}^{\text{sub}}(Q) + R_{\uparrow/\downarrow}^{+\text{SOI}}(Q)} \qquad 6$$

$R^{\text{sub}}(Q)$ represents the reflectivity for the empty MRL substrate while $R^{+\text{SOI}}(Q)$ corresponds to the reflectivity after the SOI is added to this substrate. Plotted sensitivity allow to visualize the effect of the SOI on the reflectivity curve and to gauge the information gained with the magnetic contrast. An ideal reference layer will exhibit a strong and consistent sensitivity over a broad Q-range to properly probe the SOI as well as a large difference based on the spin-state for maximum information gain to resolve the phase problem.

To evaluate the global sensitivity of the MRL across the momentum transfer range, we define three figures of merit (FOMs). The sensitivity figure of merit (SFM$_{\uparrow/\downarrow}$) quantify the total area between the $S_{\uparrow}(Q)$ or $S_{\downarrow}(Q)$ curve and the horizontal line S = 0 over the momentum transfer range Q$_{\text{min}}$ to Q$_{\text{max}}$. By taking the absolute value, $|S_{\uparrow/\downarrow}(Q)|$, both positive and negative changes for the SOI are considered. The integration bounds, Q$_{\text{min}}$ and Q$_{\text{max}}$, define the measured Q-range over which the analysis is conducted. The additional magnetic contrast figure (MCF) measures the difference between the spin-states that is a measure of the information gain from the two independent channels.

$$\text{SFM}_{\uparrow/\downarrow} = \int_{Q_{\text{min}}}^{Q_{\text{max}}} |S_{\uparrow/\downarrow}(Q)| dQ \qquad 8$$

$$\text{MCF}_{\text{diff}} = \int_{Q_{\text{min}}}^{Q_{\text{max}}} |S_{\uparrow}(Q) - S_{\downarrow}(Q)| dQ \qquad 10$$

Finally, the total sensitivity figure TSF characterizes a MRL by combining the sensitivities for multiple SOI models with the magnetic contrast:

$$\text{TSF} = \sum_{\text{types}} \left( \text{SFM}_{\uparrow}^{\text{type}} + \text{SFM}_{\downarrow}^{\text{type}} + \text{MCF}_{\text{diff}}^{\text{type}} \right) \qquad 11$$

Where "type" stands for the types "normal", "rough" and "thin" seen in Table 1. Together, these metrics provide a robust quantitative framework for evaluating the performance of the MRL in characterizing the SOI.

Thin films were deposited using ion-assisted DC magnetron sputter deposition within a high vacuum environment, maintaining a background pressure of approximately 5.6×10−5 Pa. Single crystalline Si substrates, measuring 20×20×0.5 mm³ and oriented along the 001 crystal axis, underwent thorough cleaning. The cleaning process involved successive ultrasonic baths using trichloroethylene, acetone, and isopropanol for 5 minutes each, followed by drying with nitrogen gas. Argon gas (99.999% purity) was used as the sputtering medium at a pressure of 0.51 Pa (3.8 mTorr), and the substrates were maintained at room temperature. To enhance the attraction of low energy ions during film growth, a bias voltage of -30 V was applied to the substrates, and a substrate coil current of 5 A was utilized for the deposition of all samples.

In neutron reflectivity experiments on samples deposited on silicon substrates, the native $SiO_2$ layer poses challenges due to potential chemical reactions or mixing with the sample, altering its properties. Additionally, $SiO_2$'s instability can lead to further oxidation or reactions over time, complicating reliable data acquisition. A stable capping layer is crucial to prevent such interactions and ensure accurate analysis. The chosen $Al_2O_3$ capping layer resists chemical alterations, minimizes substrate interactions, and avoids phase shift concerns by maintaining a low scattering length density (SLD), preserving the sample's integrity during measurements.[23–28]

Continuous depositions of three magnetron sources for Co, Ti, and Al were maintained, with material fluxes controlled via computer-regulated shutters. The 3-inch sputtering targets for Co and Ti had purities of 99.5%, while the 2-inch Al target also had a 99.5% purity. The magnetron discharges were operated by power-regulated supplies, where the ratio between Co and Ti was controlled by altering the power over the Ti target to achieve the desired composition. The low deposition rates were critical for ensuring precise control over layer thicknesses and roughness.

X-ray reflectivity measurements were carried out using a Malvern Panalytical Empyrean diffractometer equipped with Cu-Kα radiation and a PIXcel detector. For the incident beam, a Göbel mirror alongside a 0.5° divergence slit was utilized, while the diffracted beam optics featured a parallel beam collimator with a 0.27° collimator slit.

Polarized neutron reflectometry (PNR) experiments were conducted using the MORPHEUS instrument at the Paul Scherrer Institut, PSI (in SINQ), in Switzerland. These experiments involved directing a polarized neutron beam at the sample at small incidence angles (θ) to induce reflections at each interface before detection by a He-3 detector. PNR is particularly adept at detecting the spin-dependent scattering length density (SLD) of the sample, which provides insights into its magnetization profile. The experiments produced two distinct reflectivity curves corresponding to the two possible neutron spin states. Measurements were carried out with the samples positioned in an external magnetic field of approximately 20 mT and at angles ranging from 0° to 4° 2θ, using a neutron wavelength of 4.825 Å.

## Results & Discussion

Using simulations and FOMs the SOIs combined with the various MRLs were evaluated in terms of sensitivity over a scattering vector range.

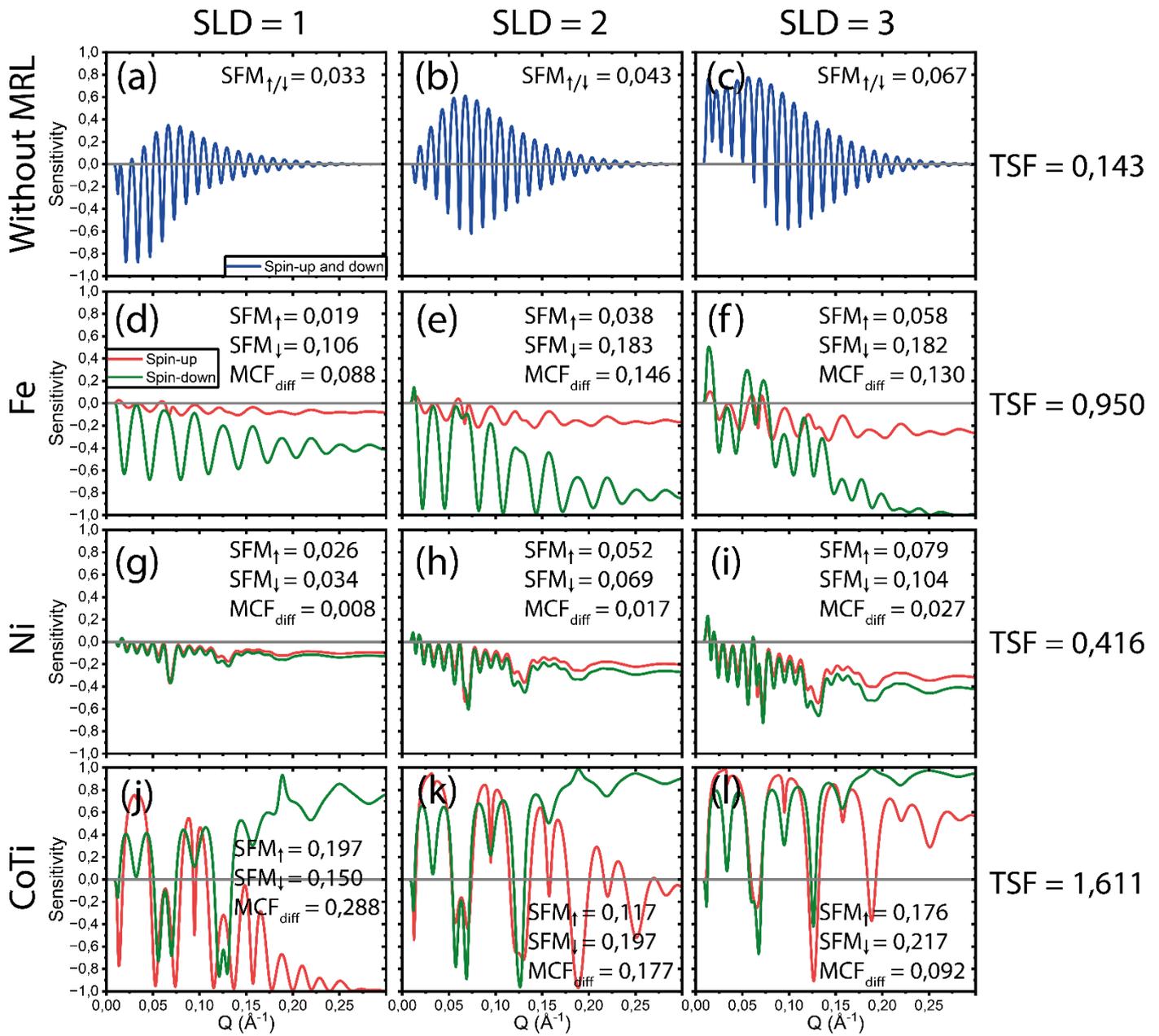

Figure 2. Sensitivity as a function over Q for "normal" SOIs with the SLDs of 1, 2 and 3 · $10^{-6}$ Å$^{-2}$ in SLD. (a-c) represents the sensitivity when not having any MRL. (d-f), (g-i) and (j-l) demonstrates the Fe, Ni and CoTi based MRLs, respectively. The total sensitivity, TSF, for each row is shown in the furthest right. Note that without MRL only is calculated for either spin-up or spin-down, thus for no-polarized neutrons should be multiplied by a factor of 2 for comparison.

Figure 2 illustrates the sensitivity as a function of Q for 12 distinct cases, all involving the "normal" SOI. For cases without a magnetic reference layer (MRL), shown in panels (a-c), polarized neutron reflectometry (PNR) is unnecessary since there is no magnetic material present. Instead, neutron reflectometry (NR) alone suffices.

Panels (d-f) display the sensitivity for PNR experiments with an Fe-based MRL. The results demonstrate a clear separation between the two spin states, attributed to the high magnetic SLD of Fe. However, the separation is predominantly observed in the spin-down sensitivity curve, which shows significant dependence on the SOI, while the spin-up curve remains relatively close to the zero as the reflectivity is dominated by the strong Fe reflectivity. Panels (h-i) show the results for a Ni-based MRL, where the magnetic SLD is lower compared to Fe. While some separation between spin-up and spin-down states is observed, it occurs mainly at higher Q-values and diminishes as the SLD of the SOI decreases.

Lastly, panels (j-l) present the sensitivity for the novel $Co_{0.64}Ti_{0.36}$ MRL. These results exhibit a pronounced difference between spin-up and spin-down sensitivity curves, with overall sensitivity levels surpassing those of both Fe- and Ni-based MRLs.

The visuals and the calculated total sensitivity TSF for all cases clearly indicate that PNR with an MRL is significantly more effective than NR without any MRL. When comparing TSF between the different MRLs, the results suggest that a higher magnetic SLD generally improves the separation between spin-up and spin-down states. However, despite Fe having a higher magnetic SLD than CoTi, the CoTi MRL demonstrates greater sensitivity for SOIs with SLDs of 1 and 2. This implies that a lower overall SLD in the MRL, combined with a clear difference in SLD between spin-up and spin-down states, enhances sensitivity. Ultimately, the CoTi MRL achieves the highest TSF among all contenders, making it the most effective configuration for distinguishing spin states across the SOI range.

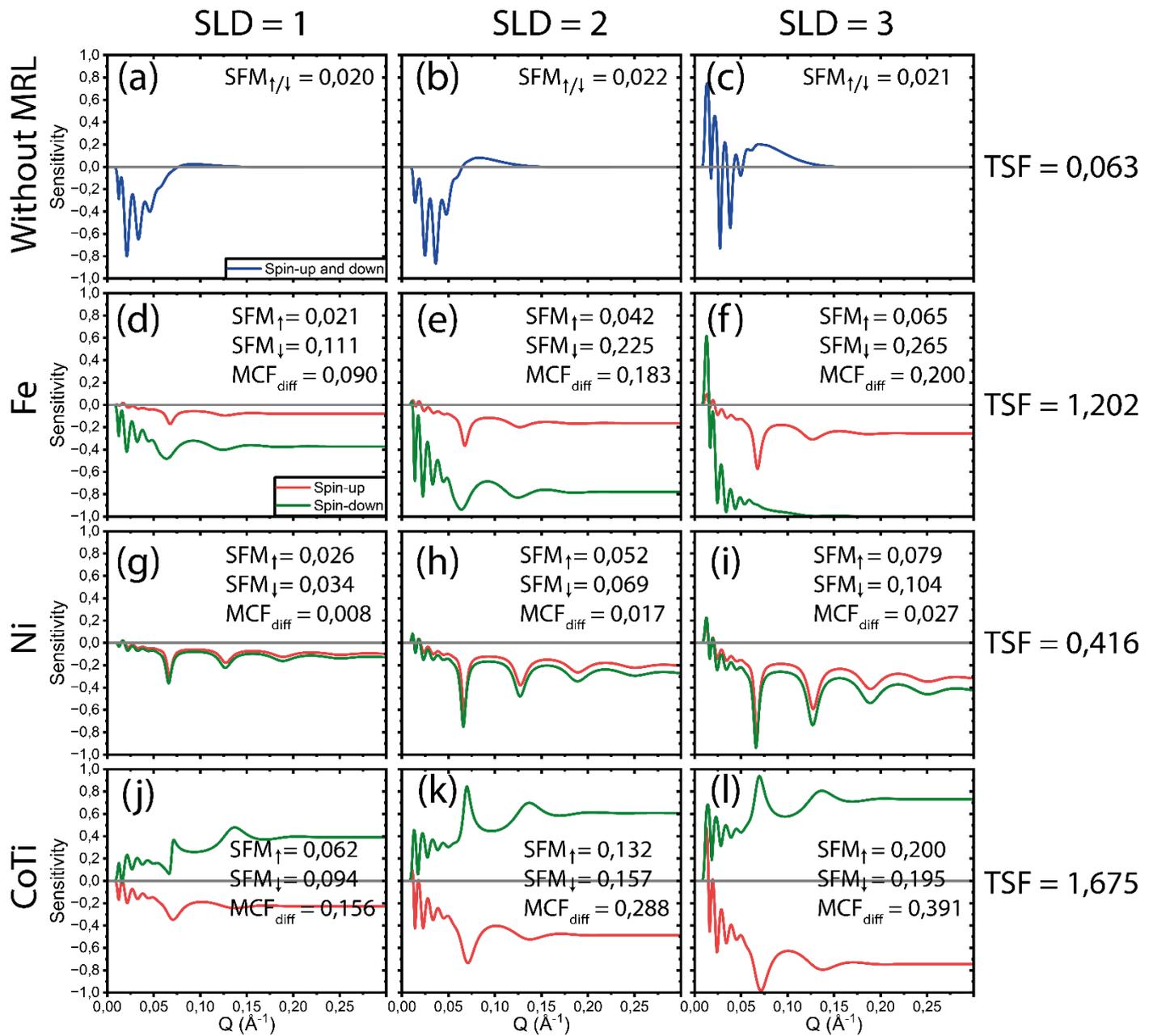

Figure 3. Sensitivity as a function over Q for "rough" SOIs with the SLDs of 1, 2 and $3 \cdot 10^{-6} \text{ Å}^{-2}$ in SLD. (a-c) represents the sensitivity when not having any MRL. (d-f), (g-i) and (j-l) demonstrates the Fe, Ni and CoTi based MRLs, respectively. The total sensitivity, TSF, for each row is shown in the furthest right. Note that without MRL only is calculated for either spin-up or spin-down, thus for no-polarized neutrons should be multiplied by a factor of 2 for comparison.

Comparing the "rough" SOIs in Figure 3 and "thin" SOIs in Figure 4 to Figure 2, similar conclusions can be drawn. In this scenario, the superior performance of the CoTi MRL becomes even more pronounced compared to the other MRLs. Additionally, it is evident that without an MRL, the sensitivity is significantly diminished, with virtually no measurable sensitivity beyond Q = 0.13 Å$^{-1}$. This highlights the critical role of incorporating an MRL in characterizing details and roughness of the SOIs that manifest at higher reflection angles.

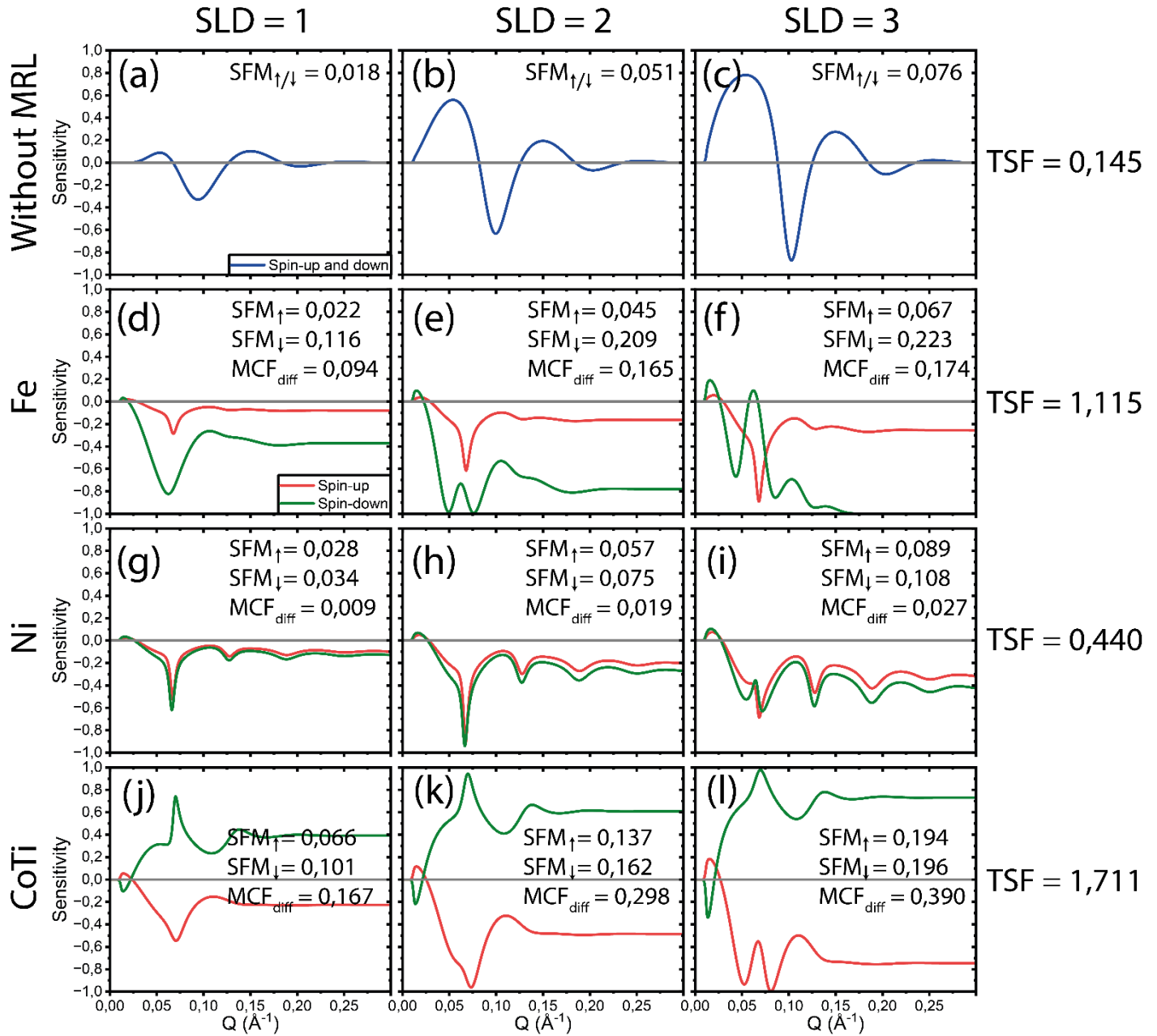

Figure 4. Sensitivity as a function over Q for "thin" SOIs with the SLDs of 1, 2 and 3 · 10$^{-6}$ Å$^{-2}$ in SLD. (a-c) represents the sensitivity when not having any MRL. (d-f), (g-i) and (j-l) demonstrates the Fe, Ni and CoTi based MRLs, respectively. The total sensitivity, TSF, for each row is shown in the furthest right. Note that without MRL only is calculated for either spin-up or spin-down, thus for no-polarized neutrons should be multiplied by a factor of 2 for comparison.

While the superior performance of CoTi is noteworthy on its own, its true strength lies in the tunability of its composition. By adjusting the ratio of Co to Ti during production, the nuclear SLD of the CoTi layer can be tailored within the range of −1.925 · 10$^{-6}$ Å$^{-2}$, as for Ti, to 2.265 · 10$^{-6}$ Å$^{-2}$, as for Co. Additionally, diluting the magnetic layer (Co) with a non-magnetic element like Ti reduces the magnetic density within the layer, thereby decreasing the magnetic SLD. This allows the magnetic SLD to be tuned from 0 to 3.9 · 10$^{-6}$ Å$^{-2}$, offering remarkable flexibility in optimizing the CoTi MRL for specific experimental needs.

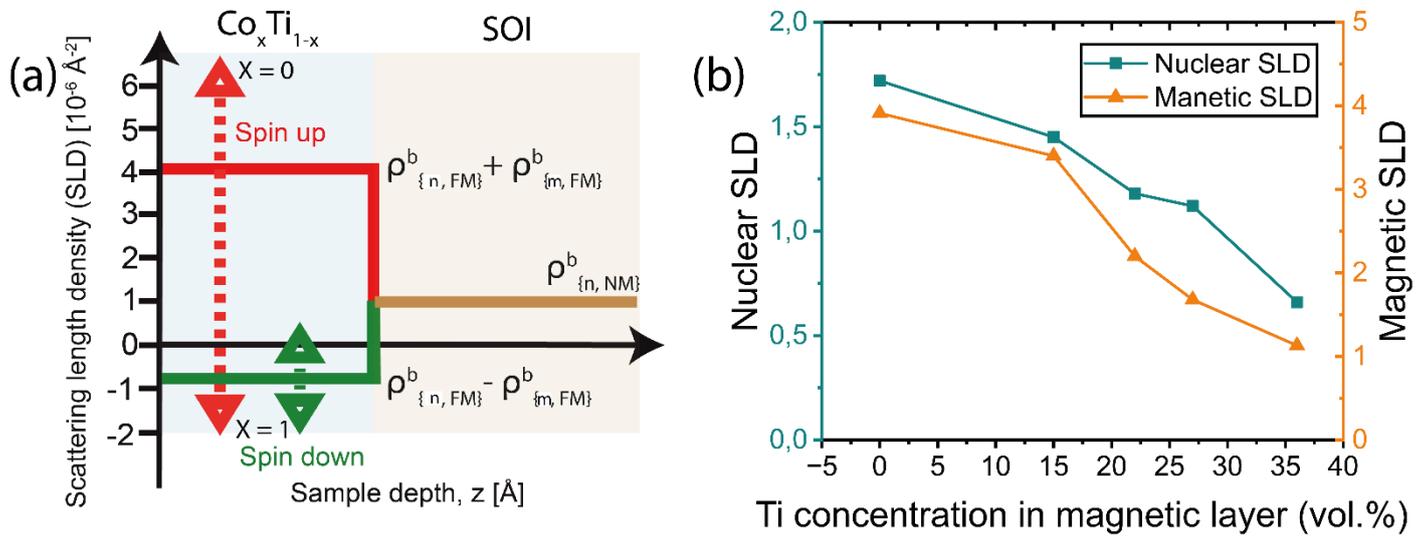

Figure 5. SLD tuning concept and fits on PNR measurements. (a) shows the theoretical SLD tuning concept where the dashed arrows red and green show tuning range for spin-up and spin-down respectively. (b) shows the fitted nuclear SLD and magnetic SLD, respectively, for 5 samples with different volumetric ratios between Co and Ti.

The tunability range of the CoTi MRL is illustrated in Figure 5(a). To validate the concept of SLD tunability, six CoTi MRL samples were fabricated, each capped with an $Al_2O_3$ layer. The corresponding measurements, fits, and SLD profiles are provided in the Supplementary Information. Figures 5(b) plot the nuclear and magnetic SLD values as a function of Ti concentration in the CoTi MRL. As expected, both the nuclear and magnetic SLDs exhibit a clear decrease with increasing Ti concentration, confirming the tunability of the SLD. This capability to precisely optimize the spin-up and spin-down SLDs within the MRL presents a powerful tool for achieving reliable and interpretable results from the SOI, making the CoTi MRL highly adaptable for a wide range of experimental conditions.

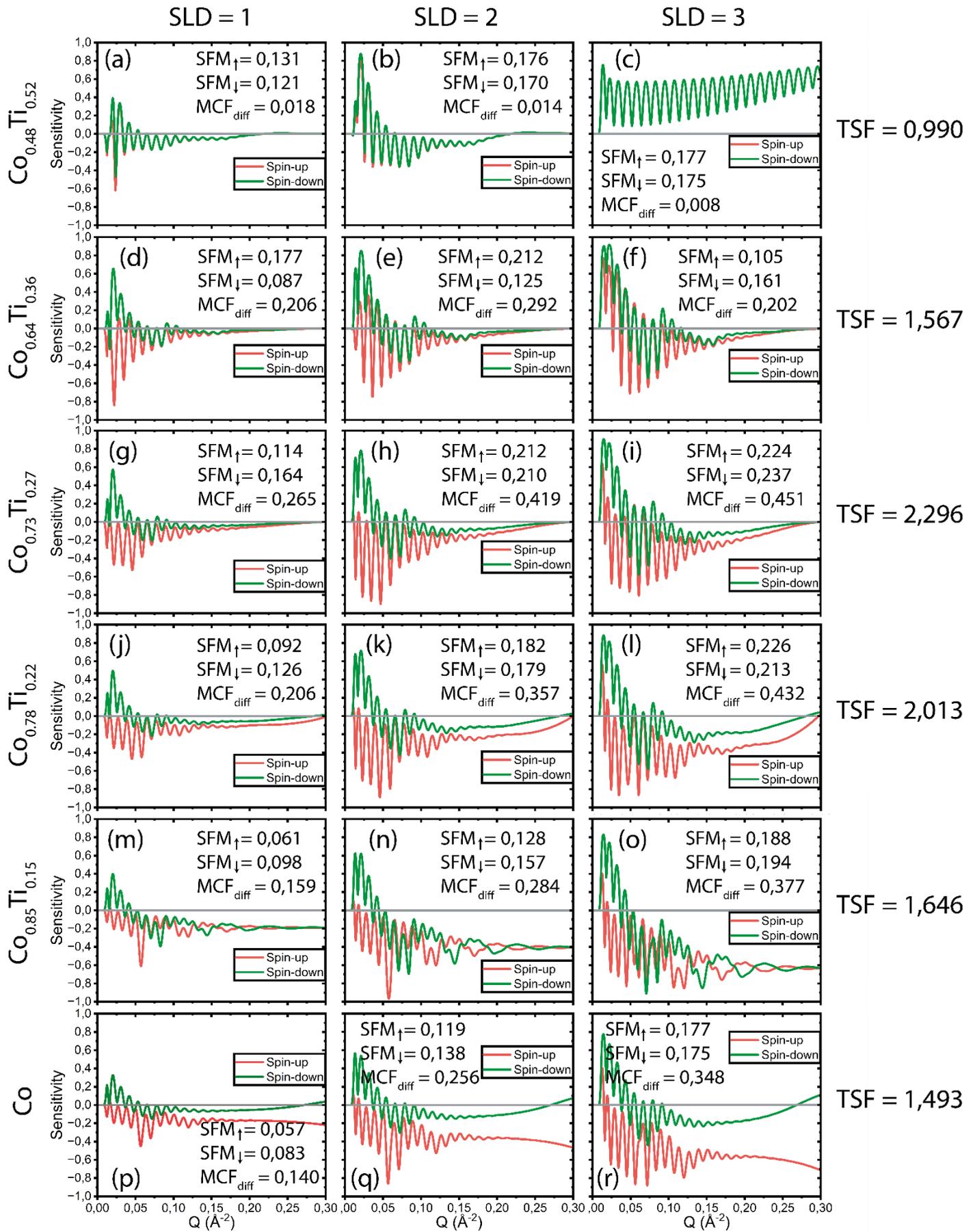

Figure 6. Sensitivity as a function over Q for "normal" SOIs with the SLDs of 1, 2 and 3 · $10^{-6}$ Å$^{-2}$ in SLD. (a-c), (d-f), (g-i), (j-l), (m-o) and (p-r) demonstrates sensitivity for 6 different CoTi based MRLs, namely $Co_{0.48}Ti_{0.52}$, $Co_{0.64}Ti_{0.36}$, $Co_{0.73}Ti_{0.27}$, $Co_{0.78}Ti_{0.22}$, $Co_{0.85}Ti_{0.15}$ and Co respectively. The total sensitivity, $TSM_{tot}$, for each row is shown in the furthest right.

Figure 6 presents the sensitivity as a function of Q for normal SOIs with SLDs of 1, 2, and $3 \cdot 10^{-6}$ Å$^{-2}$ on six different CoTi MRLs, each with varying Co-to-Ti ratios. The results reveal that MCF$_{diff}$ achieves its highest value for Co$_{0.73}$Ti$_{0.27}$ across all SOI SLDs, indicating that this ratio maximizes the difference between spin-up and spin-down states. However, when visually comparing the contrast between the spin states, the separation appears more pronounced at Q-ranges above 0.15 Å$^{-1}$ for MRLs with higher Co content, reaching its peak with pure Co due to its highest magnetic SLD. This tunability allows researchers to tailor the MRL to the specific Q-range of interest for the SOI, ensuring optimal sensitivity for features of scientific importance. Among the MRLs tested however, Co$_{0.73}$Ti$_{0.27}$ exhibits the highest TSF value, making it a reliable and versatile choice for experiments. As a result, Co$_{0.73}$Ti$_{0.27}$ emerges as a safe and effective option for most applications, balancing sensitivity and performance across varying conditions.

# Conclusion

This study provides a systematic evaluation of optimal magnetic reference layers (MRLs) and introduces a versatile framework to enhance sensitivity for studying a wide range of materials using reflectometry. CoTi-based MRLs were shown to significantly outperform Fe- and Ni-based MRLs, as well as the absence of an MRL, in enhancing the sensitivity of neutron reflectometry. Across a matrix of nine samples of interest (SOIs) varying in thickness, roughness, and scattering length density (SLD: 1, 2, $3 \cdot 10^{-6}$ Å$^{-2}$), CoTi consistently achieved the highest spin contrast (MCF$_{diff}$) and total sensitivity (TSF). The tunability of CoTi, enabled by adjusting the Co-to-Ti ratio, allows for tailored nuclear and magnetic SLDs, with Co$_{0.73}$Ti$_{0.27}$ identified as the optimal composition. This versatility ensures superior performance across diverse SOIs and Q-ranges, establishing CoTi as a robust and flexible choice for PNR experiments, particularly in soft matter and thin-film research. The ability to tailor CoTi MRLs for specific experimental requirements, combined with the open-source availability of computational tools, ensures that this approach can be readily adopted and further developed by the scientific community. These advancements have the potential to broaden the applicability of reflectometry in exploring novel materials, complex interfaces, and thin-film systems with higher sensitivity and accuracy.

# Acknowledgements

The authors thank the MORPHEUS beamline at PSI for the neutron beamtime. Hans Werthéns grant, 2022-D-03 (A.Z), Royal Academy of Sciences Physics grant, PH2022-0029 (A.Z), the Lars Hiertas Minne foundation grant FO2022-0273 (A.Z), the Längmannska Kulturfonden grant BA23-1664 (A.Z).